\documentclass[twocolumn,nofootinbib,superscriptaddress,preprintnumbers,floatfix]{revtex4}
\usepackage[bookmarks=false,hyperfootnotes=false]{hyperref}
\usepackage{graphicx}
\usepackage{amsmath}
\usepackage{color}

\newcommand{\eq}[1]{Eq.~\eqref{eq:#1}}

\newcommand{\eqss}[3]{Eqs.~\eqref{eq:#1}, \eqref{eq:#2}, and \eqref{eq:#3}}

\newcommand{\fig}[1]{Fig.~\ref{fig:#1}}

\newcommand{\nn}{\nonumber}
\newcommand{\new}{\nn\\}

\newcommand{\ord}[1]{\mathcal{O}(#1)} % order of magnitude

\newcommand{\df}{\mathrm{d}} % differential ("dx" = \df x)
\renewcommand{\ln}{\log} % make log and ln the same
\newcommand{\plus}{{\! +}}

\newcommand{\bn}{\bar n}

\newcommand{\ep}{\epsilon}

\renewcommand{\vec}[1]{\mathbf{#1}}

\newcommand{\cA}{ \mathcal{A} }
\newcommand{\cD}{ \mathcal{D} }

\newcommand{\cS}{ \mathcal{S} }

\newcommand{\as}{\alpha_s}

\newcommand{\mcdot}{\!\cdot\!}

\newcommand{\T}{\mathrm{T}} %transverse momentum
\newcommand{\kt}{k_T}
\newcommand{\alggen}[2][\alg]{\Theta_{#1}^{\, #2}}

\newcommand{\algcone}[1]{\Theta_{\rm cone}^{\, #1}}

\newcommand{\alg}{\cA}

\newcommand{\obsgen}[2][\meas]{\Delta_{#1}^{\, #2}}

\newcommand{\obs}[1]{\sigma_{#1}}
\newcommand{\obst}[1]{\widetilde{\sigma}_{#1}}
\newcommand{\obshat}[1]{{\sigma}_{#1}}
\newcommand{\obshatt}[1]{{\widetilde{\sigma}}_{#1}}
\newcommand{\meas}{{\cal M}}

\newcommand{\obsmom}[2]{\langle\obs{#1}^{#2}\rangle}
\newcommand{\obsmomt}[2]{\langle\obst{#1}^{#2}\rangle}

\newcommand{\sumpairs}[2]{\sum_{\langle #1,#2 \rangle}}

\newcommand{\dimint}[2]{\int\!\frac{\df^{#2}#1}{(2 \pi)^{#2}}}
\newcommand{\prefactor}[1]{N_{#1}}
\newcommand{\integrand}{I}

\begin{document}

%%%%%%%%%%%%%%%%%%%%%%%%%%%%%%%%%%%%%%%%%%%%%%%%%%%%%%%%%%%%%%%%%%%%%%%%%%%%%%%%
% Title page
%%%%%%%%%%%%%%%%%%%%%%%%%%%%%%%%%%%%%%%%%%%%%%%%%%%%%%%%%%%%%%%%%%

\title{Subtractions for SCET Soft Functions}

\author{Christian W.~Bauer}
\affiliation{Ernest Orlando Lawrence Berkeley National Laboratory,
University of California, Berkeley, CA 94720\vspace{0.5ex}}

\author{Nicholas Daniel Dunn}
\affiliation{Ernest Orlando Lawrence Berkeley National Laboratory,
University of California, Berkeley, CA 94720\vspace{0.5ex}}

\author{Andrew Hornig}
\affiliation{University of Washington, Seattle, WA 98195-1560, USA\vspace{0.5ex}}

%%%%%%%%%%%%%%%%%%%%%%%%%%%%%%%%%%%%%%%%%%%%%%%%%%%%%%%%%%%%%%%%%%%%%%%%%%%%%%%%
\begin{abstract}
We present a method to calculate the soft function in Soft-Collinear Effective Theory to NLO for $N$-jet events, defined with respect to arbitrarily complicated observables and algorithms, using a subtraction-based method. We show that at one loop the singularity structure of all observable/algorithm combinations can be classified as one of two types. Type I jets include jets defined with inclusive algorithms for which a jet shape is measured. Type II jets include jets found with exclusive algorithms, as well as jets for which only the direction and energy are measured. Cross sections that are inclusive over a certain region of phase space, such as the forward region at a hadron collider, are examples of Type II jets. We show that for a large class of measurements the required subtractions are already known analytically, including traditional jet shape measurements at hadron colliders. We demonstrate our method by calculating the soft functions for the case of jets defined in $\eta$-$\phi$ space with an out-of-jet $p_T$ cut and a rapidity cut on the jets, as well as for the case of 1-jettiness.
\end{abstract}
%%%%%%%%%%%%%%%%%%%%%%%%%%%%%%%%%%%%%%%%%%%%%%%%%%%%%%%%%%%%%%%%%%%%%%%%%%%%%%%%

\maketitle

%%%%%%%%%%%%%%%%%%%%%%%%%%%%%%%%%%%%%%%%%%%%%%%%%%%%%%%%%%%%%%%%%%%%%%%%%%%%%%%%
\section{Introduction}
\label{sec:intro}
%%%%%%%%%%%%%%%%%%%%%%%%%%%%%%%%%%%%%%%%%%%%%%%%%%%%%%%%%%%%%%%%%%%%%%%%%%%%%%%%
It is well known that the perturbative expansion of jet cross sections generally contains large logarithmic terms. At each order in perturbation theory, there are powers of logarithms of ratios of scales, such as the ratio of the jet mass over the jet energy or the ratio of the invariant mass between two jets over the energies of the jets. In many cases, the presence of these logarithms spoils the convergence of perturbation theory, and for that reason such large logarithms are often resummed to all orders in the perturbative expansion. 

Several techniques have been developed to allow the resummation of these logarithms~\cite{Collins:1989gx}, and recently it has been shown how soft-collinear effective theory (SCET)~\cite{Bauer:2000ew,Bauer:2000yr,Bauer:2001ct,Bauer:2001yt} can be used to resum logs. The first step in resummation is to factorize an $N$-jet cross section into hard, jet, and soft functions
\begin{eqnarray}
\df \sigma_N &\sim& B_N \, H_N\times \left[ J\ \right]^n \otimes S_N
\label{eq:SCETgeneric}
\,.\end{eqnarray}
Here $B_N$ denotes the Born-level cross section in full QCD, $H_n$ reproduces the virtual corrections of full QCD, while the $J$'s and $S_n$ together encode real emission diagrams in the collinear and soft limits. There is one jet function for each jet in the final state, and both the jet and soft functions depend on the algorithm used to define the jets, as well as the observables measured. Note that \eq{SCETgeneric} must be modified in the case of hadron collision, since there will also be PDFs and, for some measurements, beam functions \cite{Stewart:2009yx}. At tree level, the hard, jet, and soft functions are trivial, but each has to be calculated order by order in perturbation theory. 

The simplest jet definition involves exactly two jets, each consisting of all particles in one of the two hemispheres defined by a plane perpendicular to the thrust axis, and is typically only used in $e^+e^-$ collisions. In this case resummation has been achieved at  NNNLL \cite{Becher:2008cf}. 
%The hard, jet, and soft functions for this definition are known to two-loop accuracy \cite{???}, such that precision predictions can be obtained. One example is the thrust distribution in $e^+ e^-$ collisions, for which NNNLL resummation has been achieved \cite{Becher:2008cf}. 

For more complicated jet definitions, however, the required calculations are more involved, and in many cases we do not know the NLO results for the jet and soft functions. One counter example is cone or inclusive $\kt$-type algorithms in $e^+e^-$ collisions, where the distance measure is the angle $\theta$ of each particle with respect to the jet axis, and the total energy outside of the jets is less than $\Lambda$. For this example, the jet and soft functions are known to $\ord\as$~\cite{Ellis:2010rw}.

Unlike for $e^+e^-$, jet definitions at hadron colliders are usually required to be boost invariant. As a result, distance is usually measured in $\eta$-$\phi$ space, where $\eta$ and $\phi$ are defined with respect to the beam axis, and there is a restriction on the total $|\vec{p_\T}|$ outside of jets, rather than energy. Other, more inclusive jet definitions, such as $N$-jettiness \cite{Stewart:2010tn}, also have complicated dependence on the kinematics of the event, such that the NLO results for the jet and soft functions are not known. The absence of these results has been one of the biggest hurdles in deriving more precise predictions for jet cross sections at hadron colliders, and the solution to this problem is the topic of the current paper.

A generic $N$-jet cross section is defined by a jet algorithm, which identifies the regions of phase space belonging to each jet and includes restrictions on the out-of-jet radiation, and possibly one or more jet shapes, which measure functions of the final state particles in each of these jets. For an $N$-jet cross section, where an observable is measured for $m$ jets, the ${\cal O}(\alpha_s)$ contribution to the soft function can be written as
\begin{align}
\cS^{(1)}(\alg,\meas;\{\obs{m}\})&=\sumpairs{i}{j}\dimint{k}{d} \prefactor{ij}(k)\alggen{}(k)\obsgen{}(k)\new
%\end{align}
%\begin{align}
&\equiv\sumpairs{i}{j}\dimint{k}{d}\integrand_{ij}(\alg,\meas;k)\,,
\label{eq:softdef}
\end{align}
where $\prefactor{ij}$ is defined as
\begin{align}
\prefactor{ij}=-g^2\mu^{2\ep} \, \vec{T}_i \mcdot  \vec{T}_j \, \frac{n_i \cdot n_j}{n_i \cdot k \, n_j \cdot k} 2\pi\,\delta(k^2)\theta(k^0)
\,.\end{align}
 In \eq{softdef}, the sum goes over all pairs of soft emission sources $i$ and $j$, which includes the jets and possibly the beams, in the case of hadron collisions. The function $\alggen{}$ encodes the action of the jet algorithm $\alg$, $\obsgen{}$ represents the measurement $\meas$, and $\obs{i}$ is the value of the $i^{\rm th}$ jet shape. We will restrict ourselves to measurements of the form
\begin{align}
\alggen{}\obsgen{}=\sum_{k=0}^N\alggen{k}\obsgen{k}\,,
\end{align}
where $\alggen{i}$ restricts the soft gluon $k$ to be part of jet $i$, $\alggen{0}$ forces the $k$ to be outside of all jets, while still contributing to the jet cross section (typically enforcing a maximum energy or $p_T$ value).  The function $\obsgen{i}$ measures an observable in jet $i$, while setting the observables in the other jets to zero
\begin{align}
\obsgen{k}(k)=\delta\left(\obs{k}-\obshat{k}(k)\right)\, \prod_{l\neq k}\delta\left(\obs{l}\right)
\,.\end{align}
Note that for a jet where no jet shape is measured, $\obsgen{k} = \prod_l\delta(\obs{l})$.

As an example of a jet algorithm, consider a cone algorithm of size $R$ at an $e^+e^-$ collider, where the out-of-jet energy is restricted to be less than $\Lambda$. This gives
\begin{align}
\algcone{k}(k)=&\theta\left(\frac{n_k\mcdot k}{\bn_k\mcdot k} < \text{tan}^2\frac{R}{2}\right)\new
\algcone{0}(k)=&\left(1-\sum_{k=1}^N\algcone{k}(k)\right)\theta(k^0<\Lambda)\,.
\end{align}

At one loop, it can be shown that the soft function is at most $1/\ep^2$ divergent, which means that the result can be written as
\begin{align}
\label{eq:softdist}
\cS^{(1)}\left(\alg,\meas;\{\obs{m}\}\right)=&\obsmom{}{0}\prod_k\delta\left(\obs{k}\right)\\
&\hspace{-1cm}+\sum_k\left[ 4\obsmom{k}{2}-\obsmom{k}{1} \right]\left(\frac{1}{\obs{k}}\right)_\plus\prod_{l\neq k}\delta\left(\obs{l}\right)\new
&\hspace{-1cm}+\sum_k\left[ 4\obsmom{k}{2}-2\obsmom{k}{1} \right]\left(\frac{\ln\obs{k}}{\obs{k}}\right)_\plus\prod_{l\neq k}\delta\left(\obs{l}\right)\,,\nonumber 
\end{align}
where we have defined the moments of the soft function with respect to the $\obs{k}$ as
\begin{align}
\label{eq:coeff}
\obsmom{k}{n}&=\prod_{l}\int_0^1\!\!\df\obs{l}\, \obs{k}^n\,\cS^{(1)}\left(\alg,\meas;\{\obs{m}\}\right) \new
&\equiv  \dimint{k}{d}  \integrand_k^n(k)  
\,.\end{align}
Here we have assumed that all $\obs{i}$ are normalized to 1. Note that we are suppressing the dependence of the moments on the algorithm and measurement. In terms of the integral in \eq{softdef}, the zeroth moment can be written as
\begin{align}
\obsmom{}{0}=&\sumpairs{i}{j}\dimint{k}{d}\prefactor{ij}(k) \Bigg[ \sum_{k\in {\rm meas}}\alggen{k}(k)\theta(\obshat{k}(k)<1)\new
&  \qquad +\sum_{k\notin {\rm meas}}\alggen{k}(k) + \alggen{0}(k) \Bigg]
\,,\end{align}
where the subscript ``meas'' denotes the set of $m$ jets for which an observable is measured. The higher moments for measured jets ($k\in \text{meas}$) are
\begin{align}
\obsmom{k}{n>0}=&\sumpairs{i}{j}\dimint{k}{d}\prefactor{ij}(k)\obshat{k}^n(k)\alggen{k}(k)\theta(\obshat{k }(k)<1)
\,,\end{align}
while for $k\notin\text{meas}$, $\obsmom{k}{n>0}=0$.

Note that the moments $\obsmom{i}{n}$ are in general divergent, with the divergences arising from the soft ($k^0\to0$), collinear ($n_i\cdot k\to0$), and ultraviolet ($k^0\to\infty$) limits. This means that they have to be calculated analytically, with regulators for all divergences, which is in general only feasible for the simplest jet algorithms. However, consider the difference of two soft functions, $\cS(\alg,\meas;\{\obs{m}\})$  and $\widetilde{\cS}(\widetilde{\alg},\widetilde{\meas};\{\obst{m}\})$, which are defined with different algorithms and measurements. We will look at the difference
\begin{align}
\cD_k^n 
& = \obsmom{k}{n}-\obsmomt{k}{n}\,,
\label{eq:diff}
\end{align}
where $\obsmom{i}{n}$ are moments of $\cS^{(1)}$ and $\obsmomt{i}{n}$ are moments of $\widetilde{\cS}^{(1)}$. Note that in practice it is often useful to define the moments $\sigma^n$ and differences $\cD^n$ for each color structure separately, which we will denote by $[\sigma^n]_{ij}$ and $[\cD^n]_{ij}$.

In the soft limit, $k^0\to0$, IR safety dictates that $\obshat{i}(k)\to0$. This implies that all higher moments (and therefore their differences) vanish. For the zeroth moment, the observable theta functions are trivially satisfied, such that
\begin{align}
\lim_{k^0\to0}\big ( \integrand^0(k)  - \tilde \integrand^0(k)\big) \propto\sum_{k=0}^N\Bigl[\alggen{k}(k)-\alggen[\widetilde{\alg}]{p}(k)\Bigr]=0\,.
\end{align}
Here we have used the fact that gluon will not be vetoed as its energy goes to 0. This implies that it is assigned either to be in a jet or to the out-of-jet region, such that $\sum_{i=0}^N\alggen{i} = 1$ for any algorithm.

In the collinear limit, $n_i\mcdot k\to 0$, IR safety once again tells us that $\obshat{i}(k)\to0$. In addition, for fixed $k^0$, any IR-safe algorithm will assign the emission to jet $i$ in this limit, such that $\alggen{k}=\delta_{ik}$. This leads to the desired result
\begin{align}
\lim_{n_i\cdot k\to 0}\big ( \integrand^0(k)  - \tilde \integrand^0(k)\big)  \propto\sum_{k=0}^N\Bigl[\alggen{k}(k)-\alggen[\widetilde{\alg}]{k}(k)\Bigr]=0
\,.\end{align}

The final limit, $k^0\to\infty$, requires a more careful analysis. As the energy of the gluon goes to infinity, any out-of-jet cut will veto the emission, which means we only need to consider radiation in the jets. The difference of the zeroth moments is 
\begin{align}
\label{eq:UVlimit1}
\lim_{k^0\to\infty}\big ( \integrand^0(k)  - \tilde \integrand^0(k)\big)\propto&\sum_{k\notin {\rm meas}} \left[ \alggen{k}(k) - \alggen[\widetilde{\alg}]{k}(k) \right]  \\
&
\hspace{-3.6cm}+ \sum_{k\in {\rm meas}} \left[ \alggen{k}(k)\theta(\obshat{k}(k)<1) -\alggen[\widetilde{\alg}]{k}(k)\theta(\obshatt{k}(k)<1) \right]\nonumber\,,
\end{align}
while for the higher moments we find
\begin{align}
\label{eq:UVlimit2}
\lim_{k^0\to\infty}\big (\integrand_k^{n}(k)  - \tilde\integrand_k^{n}(k)\big)\propto\,&\obshat{k}^n(k)\alggen{l}(k)\theta(\obshat{k}(k)<1)\\
&-\obshatt{k}^n(k)\alggen[\widetilde{\alg}]{k}(k)\theta(\obshatt{k}(k)<1)\nonumber \,.
\end{align}

In contrast to the soft and collinear limits, where observable dependence is trivial, the UV divergences are naively sensitive to both the algorithm and observable definitions. However, we find that there are two possible cases for jets: Type I, where the bound on the observable is more restrictive
\begin{align}
{\rm Type \, I}: \lim_{k^0\to\infty}\alggen{l}(k)\theta(\obshat{l}(k)<1)=\lim_{k^0\to\infty} \theta(\obshat{l}(k)<1)
\,,\end{align}
and Type II, where the jet algorithm is more restrictive
\begin{align}
{\rm Type \, II}: \lim_{k^0\to\infty}\alggen{l}(k)\theta(\obshat{l}(k)<1)=\lim_{k^0\to\infty}\alggen{l}(k)
\,.\end{align}
Here we have defined the limit of the algorithm restriction and the observable in the UV. For jets where an observable is not measured, the jet algorithm is, by definition, more restrictive, such that these jets are always Type II. 

Clearly, if $\cS$ and $\widetilde{\cS}$ do not agree on which jets are Type I versus Type II, the difference of moments will not go to 0 and $\cD$ will be divergent as $k^0\to\infty$. Assuming they agree, \eq{UVlimit1} can be simplified, giving
 \begin{align}
&\lim_{k^0\to\infty}\big ( \integrand^0(k)  - \tilde \integrand^0(k)\big)\propto\new
&\qquad \lim_{k^0\to\infty} \sum_{p\in\text{Type I}}\Bigl[\theta(\obshat{p}(k)<1)-\theta(\obshatt{p}(k)<1)\Bigr]\new
& \qquad\hspace{-.5cm}+\lim_{k^0\to\infty}\sum_{p\in\text{Type II}}\Bigl(\alggen{p}(k)-\alggen[\widetilde{\alg}]{p}(k)\Bigr) 
\,,\end{align}
while \eq{UVlimit2} gives
\begin{align}
\label{eq:UVfinal}
&\lim_{k^0\to\infty}\big (\integrand_{k\in\text{Type I}}^{n}  - \tilde\integrand_{k\in\text{Type I}}^{n}\big)\propto \\
& \qquad\,\lim_{k^0\to\infty} \bigg[\obshat{k}^n(k)\theta(\obshat{k}(k)<1) -\obshatt{k}^n(k)\theta(\obshatt{k}(k)<1)\bigg]\nonumber \,.
\end{align}
Here we have used the fact that, for Type II jets, all higher moments are 0 \cite{long}. We now see that, as $k_0 \to \infty$, if $\lim_{k_0\to\infty}\obshat{k} = \lim_{k_0\to\infty}\obshatt{k}$ for all Type I jets and $\lim_{k_0\to\infty}\alggen{k}=\lim_{k_0\to\infty}\alggen[\widetilde{\alg}]{k}$ for all Type II jets, the integrand vanishes as $k^0\to\infty$.

These results can be used to calculate a desired soft function numerically, given an analytically known soft function. Combining \eqss{softdist}{coeff}{diff} gives
\begin{align}
\cS^{(1)}(\alg,\meas;\{\obs{m}\})=\,\widetilde{\cS}^{(1)}(\widetilde{\alg},\widetilde{\meas};\{\obst{m}\to\obs{m}\})\new
+\cD^0\prod_i\delta(\obs{i}) + \sum_i \cD_i^1\prod_{j\neq i}\delta\left(\obs{j}\right)\left(\frac{1}{\obs{i}}\right)_\plus
\,,\end{align}
where we have used that the $(\log \obs{i} /\obs{i})_+$ distribution must agree \cite{long}. The notation $\obst{m}\to\obs{m}$ means that, in $\widetilde{\cS}^{(1)}$, all $\obst{i}$ are to be replaced by $\obs{i}$ for Type I jets, while for Type II jets, $\prod_{i\notin \widetilde{\text{meas}}}\delta(\obst{i})$ becomes $\prod_{i\notin\text{meas}}\delta(\obs{i})$. As long as the subtraction soft function has the same UV observable dependence as the target for all Type I jets, as well as the same UV algorithm dependence for all Type II jets, the $\cD_i^n$ are finite and can be calculated numerically. If $\widetilde{\cS}$ is known analytically, the result for $\cS$ can be fully computed at one loop. We will now illustrate this procedure with several examples. For a longer discussion, see \cite{long}.

For our subtraction function, we will use the results from \cite{Ellis:2010rw}. By combining the pieces of the soft function calculated therein, we can construct a subtraction for any soft function at NLO that meets the following two conditions: observables for Type I jets must be symmetric about the jet axis and behave like an angularity \cite{Berger:2003iw} with $a < 1$ in the UV, and all Type II jets must be found using either a cone or inclusive $\kt$-type algorithm, with the $\theta$-angle relative the jet axis as a measure.
\begin{figure}[htbp]
\includegraphics[width=1\columnwidth]{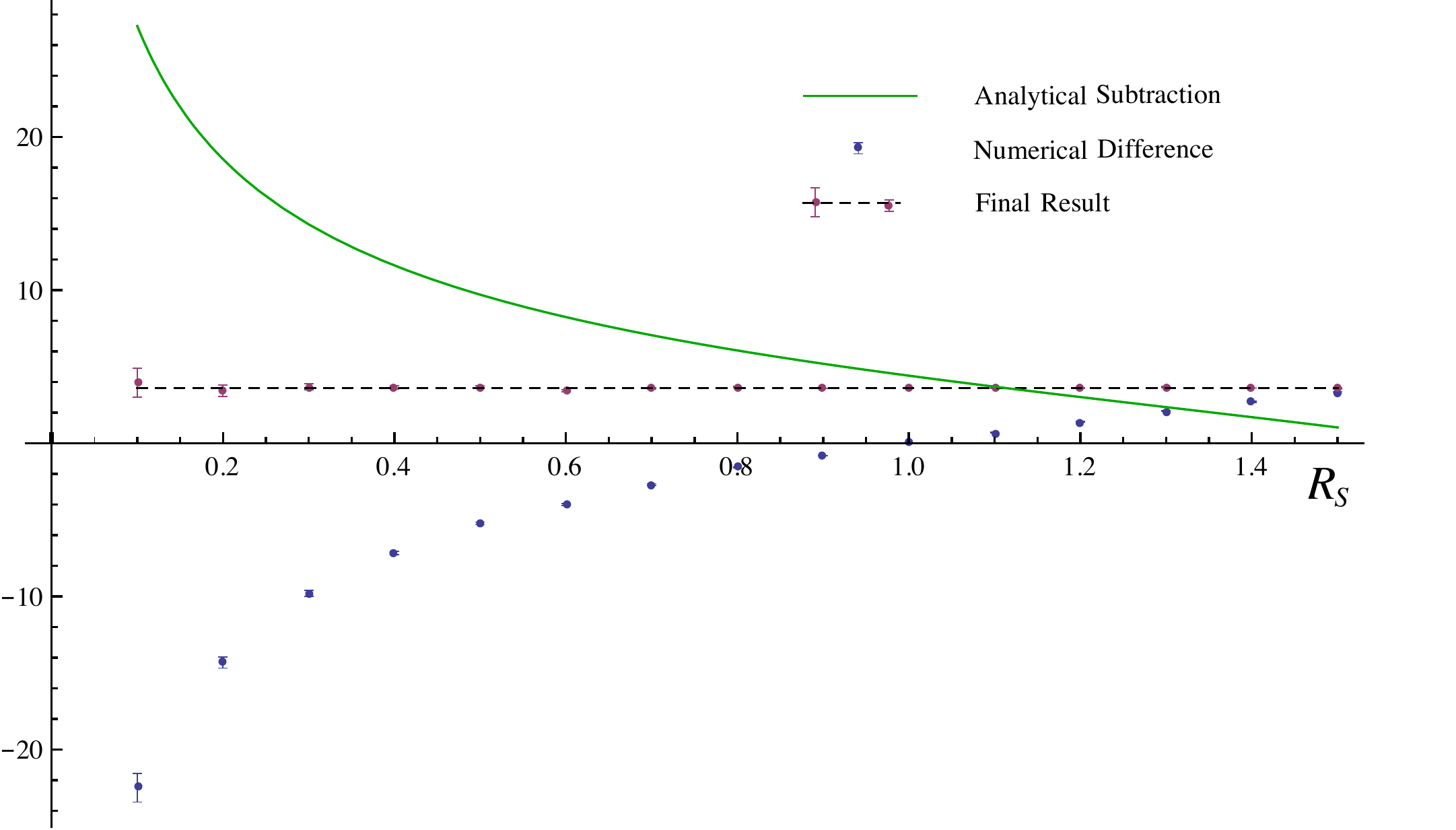}
\caption{The coefficient $[\cD^0]_{ij}$ normalized to $\frac{\as}{\pi} \vec{T}_i\mcdot\vec{T}_j$ for three equally-spaced jets as a function of the subtraction jet size $R_S$. 
\label{fig:RtoRp}
 }
\end{figure}

Using the notation of \cite{Ellis:2010rw}, our subtraction function is
\begin{align}
\widetilde{\cS}^{(1)}=\sumpairs{i}{j}\Biggl[&\Bigl(S^{\text{incl}}_{ij}+\sum_k^NS_{ij}^k\Bigr)\prod_{l}^{\widetilde{m}}\delta(\tau_a^l)\new
&+\sum_{k}^{\widetilde{m}}S^{\text{meas}}_{ij}(\tau_a^k)\prod_{l\neq k}^{\widetilde{m}}\delta(\tau_a^l)\Biggr]\,.
\end{align}
Here, $\widetilde{m}$ is set to be the number of Type I jets in $\cS$, while the $\tau_a^k$ are chosen such that $\lim_{k^0\to\infty}(\obs{k}-\tau_a^k) = 0$ for all Type I jets.

As an initial check of our method, we seek to reproduce the known analytical result for $\cD^0$ of three equally separated jets, each with energy 125 GeV, found using a cone algorithm of size $R$ measured as the angle with respect to the jet and an out-of-jet energy cut of $\Lambda = 50$ GeV. Jet thrust ($a=0$) will be measured on each jet. Our subtraction will use size $R_S$, the same out-of-jet energy cut, and $\mu=\Lambda=50$ GeV. As we will show in more detail in~\cite{long}, the result reduces to a simple numerical integral over the solid angle. We see in \fig{RtoRp} that the known result is reproduced, within numerical error due to Monte Carlo integration. 

\begin{figure}[htbp]
\includegraphics[width=0.9\columnwidth]{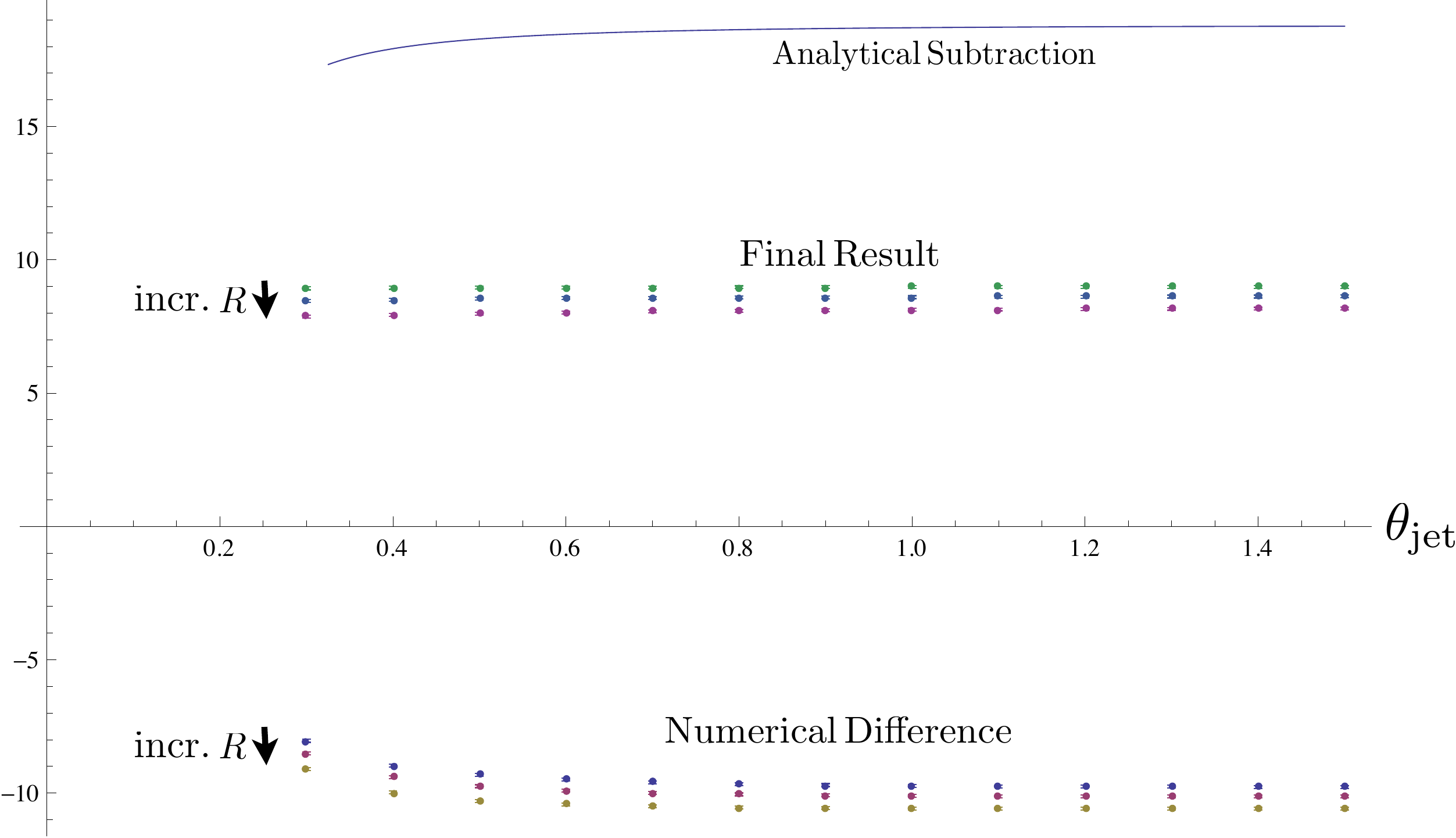}
\caption{The soft function for an $\eta$-$\phi$ jet algorithm as a function of the angle of the measured jet with different $R$ for the case when the emitters $\langle ij \rangle$ are the two beam directions. 
\label{fig:plot2}
 }
\end{figure}

In \fig{plot2}, we show the result of of the soft function for a single measured jet with $a=0$ at a hadron collider. The jet is defined with an $\eta$-$\phi$ algorithm, and has energy 125 GeV. For simplicity we will choose the beams to have energy of 125 GeV as well. To veto other jets, we use a $p_\T$ cut of 20 GeV for $|\eta| < 5$, while remaining inclusive for $|\eta| > 5$. We show three separate curves for $R = 0.4,0.7,1.0$. The plot is shown as a function of the angle of the jet with respect to the beams. Note that to remain inclusive for $|\eta| > 5$  requires the use of Type II jets in the subtraction.

\hspace{-0.5cm}

\begin{figure}[htbp]
\includegraphics[width=0.9\columnwidth]{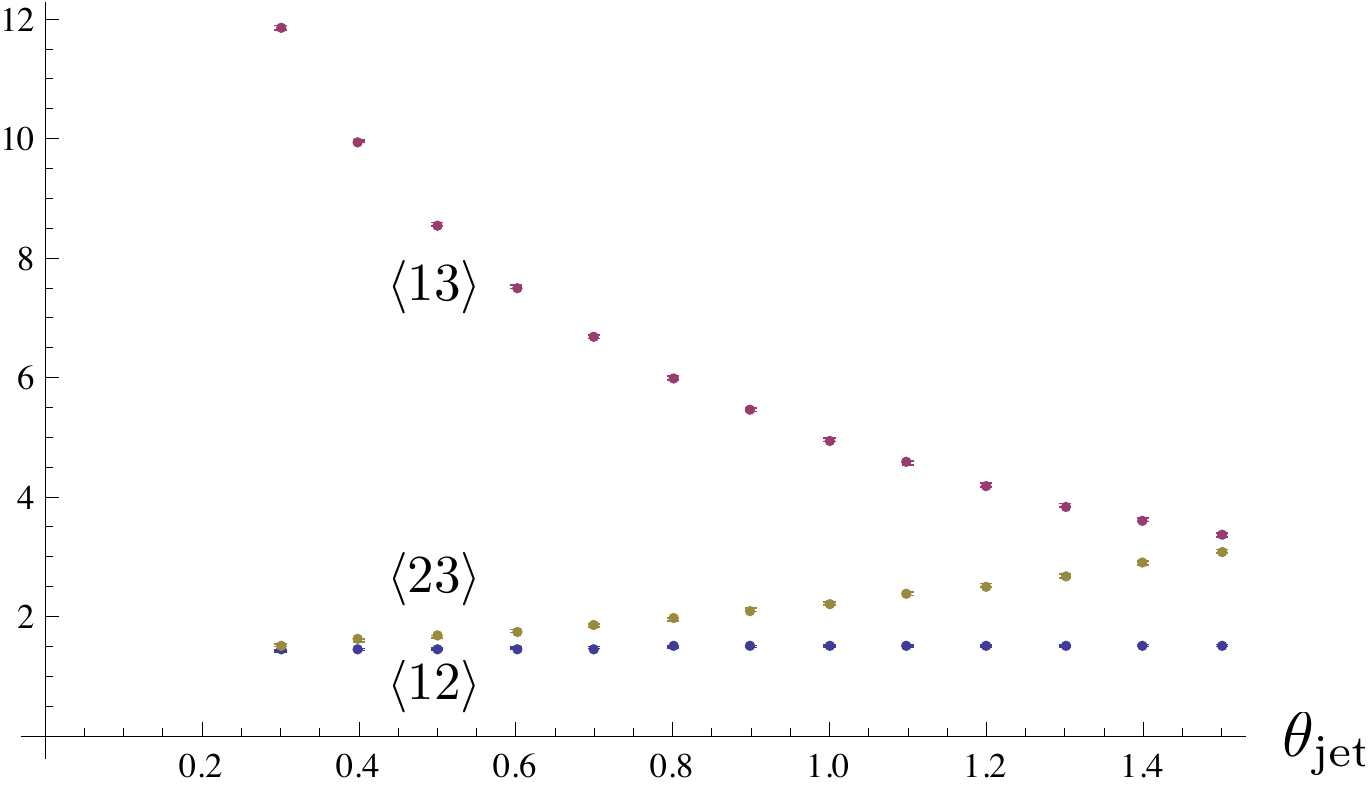}
\caption{The coefficient $\cD^0$ for one-jettiness, shown as a function of the angle of the measured jet. The labels refer to the three $\langle ij \rangle$ emitters, with 1 and 2 referring to the beams and 3 to the jet.
\label{fig:plot3}
 }
\end{figure}

Using the same kinematic configuration as in the $\eta$-$\phi$ example (two fixed back-to-back directions, with a third direction of varying angle), we show  in \fig{plot3} the result for 1-jettiness \cite{Stewart:2010tn}. Since there is a measurement in each region, only Type I jets are used in the subtraction. 

We have seen that, at one loop, subtractions can be constructed that allow for new soft functions to be calculated numerically, using previously derived analytical results. Using only calculations that exist in the literature, the soft function can be calculated for a number of phenomenologically relevant observables and algorithms, including jet shapes found using $\eta$-$\phi$ algorithms and $N$-jettiness, for arbitrary $N$. In order to calculate the soft function for most current measurements at hardon colliders, both Type I and Type II jets are required. We should also note that a similar argument can be used to calculate jet functions in the presence of algorithms; however, this is generally less relevant as algorithm corrections tend to be power suppressed. 

\hspace{-0.3cm}
%%%%%%%%%%%%%%%%%%%%%%%%%%%%%%%%%%%%%%%%%%%%%%%%%%%%%%%%%%%%%%%%%%%%%%%%%%%%%%%%
\acknowledgments
%%%%%%%%%%%%%%%%%%%%%%%%%%%%%%%%%%%%%%%%%%%%%%%%%%%%%%%%%%%%%%%%%%%%%%%%%%%%%%%%

This work was supported in part by the Director, Office of Science, Offices
of High Energy and Nuclear Physics of the U.S.\ Department of Energy under the
Contracts DE-AC02-05CH11231. AH was supported by the DOE under contract DE-FGO3-96-ER40956.

\bibliography{master_bib}
\end{document}